# Layer-dependent field-free switching of Néel vector in a van der Waals antiferromagnet

Haoran Guo[1†], Zhongchong Lin[1,2†], Jinhao Lu[1], Chao Yun[3,1*], Guanghui Han[4], Shoutong Sun[1], Yu Wu[1], Wenyun Yang[1], Dongdong Xiao[5], Zhifeng Zhu[6], Licong Peng[4,7], Yu Ye[1,7], Yanglong Hou[3,4,7*], Jinbo Yang[1,7*], Zhaochu Luo[1,7*]

[1]State Key Laboratory for Mesoscopic Physics, School of Physics, Peking University, Beijing 100871, China.
[2]Fujian Provincial Key Laboratory of Quantum Manipulation and New Energy Materials, College of Physics and Energy, Fujian Normal University, Fuzhou 350117, China.
[3]School of Materials, Sun Yat-Sen University, Shenzhen 518107, China.
[4]School of Materials Science and Engineering, Peking University, Beijing 100871, China.
[5]Beijing National Laboratory for Condensed Matter Physics, Institute of Physics, Chinese Academy of Sciences, Beijing 100190, China.
[6]School of Information Science and Technology, ShanghaiTech University, Shanghai 201210, China.
[7]Beijing Key Laboratory for Magnetoelectric Materials and Devices, Beijing 100871, China.
*Correspondence to: zhaochu.luo@pku.edu.cn (Z.L.); jbyang@pku.edu.cn (J.Y.); hou@pku.edu.cn (Y.H.); yunch@mail.sysu.edu.cn (C.Y.).
†These authors contributed equally: Haoran Guo, Zhongchong Lin.

**Abstract:**

**Two-dimensional antiferromagnets, combining the dual advantages of van der Waals (vdW) and antiferromagnetic materials, provide an unprecedented platform for exploring emergent spin-related phenomena. However, electrical manipulation of Néel vectors in vdW antiferromagnets—the cornerstone of antiferromagnetic spintronics—remains challenging. Here, we report layer-dependent electrical switching of the Néel vector in an A-type vdW antiferromagnet (Fe,Co)$_3$GaTe$_2$ (FCGT) with perpendicular magnetic anisotropy. The Néel vector of FCGT with odd-number vdW layers can be 180° reversed via spin-orbit torques. Furthermore, we achieve field-free switching in an all-vdW, all-antiferromagnet heterostructure of FCGT/CrSBr in which the noncollinear interfacial spin texture breaks the mirror symmetry. Our results establish layer-controlled spin symmetries and interfacial spin engineering as universal paradigms for manipulating antiferromagnetic order, paving the way for realising reliable and efficient vdW antiferromagnetic devices.**

**Introduction**

Owing to the internally compensated spin sublattices, antiferromagnetic materials produce negligible stray fields and are immune to disturbing magnetic fields, promising for building ultra-dense and reliable magnetic memory devices[1-3]. Besides, the THz spin dynamics resulting from antiparallel exchange-coupled spin sublattices, allows antiferromagnetic devices operating at an ultra-fast speed beyond its ferromagnetic counterpart[4-6]. One of the key capabilities for antiferromagnetic spintronics is to manipulate the antiferromagnetic order via electrical means. Pioneering works have demonstrated the capability of electrical 90° and 120° switching of Néel vectors in various thin-film antiferromagnets (*e.g.*, CuMnAs[5,7,8], Mn$_2$Au[9,10], NiO[6,11,12], $\alpha$-Fe$_2$O$_3$[13] and PtMn[14]) via staggering spin-orbit torques (SOTs) and spin Hall effect-induced SOTs that have emerged as an efficient means of manipulating the magnetic state. The 90° and 120° switching of Néel vector usually requires altering current directions between multiple writing channels, leading to the complexity of writing signal control and difficulties in device integration. By contrast, the 180° switching of antiferromagnetic order can be achieved by reversing the current polarity in a single writing channel[15-18], analogous to the SOT switching of a ferromagnet, which thus is compatible with readily ferromagnet-based SOT-magnetic random-access memory (MRAM) architecture.

In this switching scheme, the presence of a small net magnetic moment **δ*m*** that breaks the time-reversal symmetry, is crucial to realise the deterministic switching and its orientation relationship with the Néel vector decides the switching polarity. Despite several attempts at electrical 180° switching in thin-film collinear antiferromagnet[15] as well as unconventional altermagnet[16,17], the origin of the net magnetic moment in thin-film antiferromagnets is usually elusive, possibly due to Dzyaloshinskii-Moriya interaction (DMI)-induced spin canting[16-21] or uncompensated magnetic moments at defects and interfaces due to surface reconstruction[15,16,22]. DMI-induced net magnetic moments usually give the orthogonal relation with the Néel vector, *i.e.* **δ*m*** ⊥ ***n*** (Fig. 1a), while interface-induced net magnetic moments are parallel (antiparallel) to the Néel vector, *i.e.* **δ*m***//***n*** (Fig. 1b). The ambiguous origin of the net magnetic moment and its orientation relation with the Néel vector will cause the reliability issue of SOT switching in antiferromagnets. Moreover, the 180° switching of Néel vector encounters a similar challenge to the SOT switching in ferromagnets, which is the necessity for an external in-plane magnetic field, thus hindering the scalability of antiferromagnetic SOT-MRAM.

The discovery of van der Waals (vdW) magnetic materials has established an unprecedented platform for the research of emergent spin-related phenomena[23,24]. In particular, owing to their intrinsically clean surface without dangling bonds, these cleavable magnetic materials hold the potential to tailor the spin symmetries and to construct superior heterostructures at the atomic scale. In this work, we synthesize a metallic A-type vdW antiferromagnetic crystal $(Fe,Co)_3GaTe_2$ (FCGT) with perpendicular magnetic anisotropy (PMA), and fabricate the Pt/FCGT heterostructure serving as the experimental system to study SOT switching mechanism in antiferromagnets. The intrinsic net magnetic moment **δ*m***//***n*** in Pt/FCGT can be precisely modulated by the even/odd parity of vdW layers (Fig. 1c). Combining with the theoretical macrospin model, we reveal a layer-dependent parity effect that governs the switching dynamics and experimentally demonstrate that the Néel vector of FCGT with odd-number vdW layers can be 180° switched via current-induced SOTs. In addition, by constructing an all-vdW, all-antiferromagnet heterostructure of FCGT/CrSBr with noncollinear interfacial spin textures which induces the necessary mirror symmetry-breaking to realise deterministic switching, we achieve field-free 180° switching of Néel vector when applying currents along the in-plane easy axis of the antiferromagnet semiconductor CrSBr.

**Crystal growth and basic characterization**

The plate-like FCGT single crystals with sizes of about 2 × 2 × 0.02 mm are grown by the self-flux method (details seen in Methods). The crystalline structure of FCGT with $P6_3/mmc$ space group is shown in Fig. 1d, where the $(Fe,Co)_3Ga$ slabs are sandwiched by two Te layers[25-27]. According to the energy dispersive X-ray spectroscopy (EDS) results, the ratio of Fe and Co atoms is estimated to be ~4: 1 (Fig. S1). The high-resolution transmission electron microscopy (HR-TEM) image confirms the layered atomic structure of FCGT with a vdW gap is ~0.3 nm (Fig. 1e). The X-ray diffraction (XRD) pattern of the FCGT bulk presented in Fig. 1f shows only (0 0 *l*) peaks, which reveals the high crystallinity and crystal surface being the crystallographic ***ab***-plane. In addition, the temperature-dependent magnetization is measured with the magnetic field applied along the out-of-plane (***H***//***c***) and in-plane (***H***//***ab***) directions, using zero-field-cooling (ZFC) and field-cooling (FC) protocols, indicating a Néel temperature of 196 K (Fig. S2). The distinct magnetic hysteresis loops with out-of-plane and in-plane magnetic fields show that FCGT has a good PMA. Particularly, we observe a metamagnetic transition when $H//c$: the two spin sublattices start to snap at ~10 kOe and saturates to the same direction at ~23 kOe at 100 K. By contrast, when $H//ab$, the two spin sublattices exhibits a continuous canting along the direction of magnetic fields, confirming the PMA in FCGT. We then obtain the phase diagram of spin alignment at different temperatures and out-of-plane magnetic fields (Fig. 1g) and quantify the magnetic parameters of interlayer exchange coupling $J_{AF}$ and magnetic anisotropy energy $K_u$ by fitting magnetic hysteresis loops with the two-spin macrospin model[28] (Supplementary Note 2).

**Intrinsic net magnetic moment in Pt/FCGT**

We mechanically exfoliate the FCGT crystal into nm-thick flakes using Scotch tapes. The FCGT flakes with a thickness of 5-10 nm and lateral dimension of ~10 μm are then transferred onto a 200 nm-thick $SiO_2$ layer on a Si substrate in the glovebox enclosure with inert argon atmosphere. We sputter a 7 nm-thick Pt layer on the exfoliated FCGT flakes to create the Pt/FCGT heterostructure that is further patterned into Hall devices (Fig. 2a). The top Pt layer can serve as the protection layer preventing the oxidation of FCGT as well as the source of spin current due to its strong spin-orbit coupling. In the FCGT flake, the net magnetic moment is $\boldsymbol{\delta m} = \sum_{i=1}^{N} \boldsymbol{m_i}$ and the Néel vector is defined as: $\boldsymbol{n} = \boldsymbol{m_1} - \boldsymbol{m_2}$, where $\boldsymbol{m_i}$ is the unit magnetic moment of the $i^{th}$ layer from the top surface and $N$ is the number of vdW layers (shown in the inset of Fig. 2a).

To study the magnetic properties of the Pt/FCGT heterostructure, we perform the anomalous Hall effect (AHE) measurement with a relatively low current of 1 mA. The AHE hysteresis loops in Pt/FCGT devices are similar to that of the FCGT crystal (Figs. 2b and 2c), showing that the antiferromagnetism has been well preserved during the nano-fabrication process. Interestingly, we observe a sizable Hall resistance remanence $\delta R_H$ that vanishes concurrently at the Néel temperature in some Pt/FCGT heterostructures (Fig. S4). We attribute the presence of $\delta R_H$ to the uncompensated magnetic moment in the FCGT flake with odd-number vdW layers. To verify this, we quantify the vdW layer numbers in the Pt/FCGT devices with and without $\delta R_H$ by using HR-TEM. As shown in Figs. 2d and 2e, there is a sharp and clean interface between the Pt and FCGT layers. The vdW layer numbers of the FCGT whose AHE hysteresis loops are shown in Figs. 2b and 2c, are 9 and 12, respectively. We further estimate the ratio of $\boldsymbol{\delta m}/\sum_{i=1}^{N}|\boldsymbol{m_i}|$ to be ~11% in a 9-layer FCGT flake, which agrees well with the magnitude ratio (~10.1%) of $\delta R_H$ and the saturated Hall resistance $\Delta R_H$ in the Pt/FCGT device. Besides, as shown in Fig. 2g, the net magnetic moment exhibits a high anisotropy field, implying the strong magnetic coupling with the Néel vector in FCGT.

The presence of nonzero net magnetic moments that breaks the time-reversal symmetry, plays a crucial role on the realisation of electrical 180° switching and detection of Néel vectors in collinear antiferromagnets. Notably, previous reports on net magnetic moments in antiferromagnets are mainly resulted from spin canting due to DMI[16-21] and uncompensated magnetic moments at defects or interfaces[15,16,22]. Specifically, the occurrence of net magnetic moment that purely originates from DMI, requires strict spin and structural symmetries, which is rare among antiferromagnets[19,20]. In most antiferromagnets, the net magnetic moments are associated with defects or interfaces[15,16,22], which are extrinsic. Its magnitude and orientation relation with Néel vector strongly depends on the material synthesize and fabrication process. In contrast to conventional thin-film antiferromagnets, we demonstrate that the magnitude of $\boldsymbol{\delta m}$ in the vdW antiferromagnet can be precisely modulated by varying the vdW-layer number and the direction of $\boldsymbol{\delta m}$ is unambiguously parallel to the Néel vector, *i.e.* $\boldsymbol{\delta m}//\boldsymbol{n}$ (Supplementary Note 4). This controllable and intrinsic net magnetic moment can further serve as a handle for reliable SOT switching of Néel vector.

**Electrical 180° switching of Néel vector in Pt/FCGT**

We next demonstrate the electrical manipulation of Néel vector via SOTs. As shown in Fig. 3a, we employ the switching scheme used for SOT switching of a PMA ferromagnet: an in-plane magnetic field is applied along the direction of electric currents to assist the switching[29,30]. In the experiment, the FCGT flake with 11 vdW layers is used (Fig. 3b). By sweeping the currents with a fixed in-plane magnetic field, we observe a switching of the Hall resistance between the two resistance states corresponding to magnetic states of $\boldsymbol{\delta m}$ pointing to ↑ and ↓ (Fig. 3c). As $\boldsymbol{\delta m}$ is parallel to the Néel vector, the current-induced Hall resistance switching indicates the electrical switching of Néel vector. Note that the magnitude of the current-induced Hall resistance switching is almost equal to that switched by magnetic fields, showing a nearly complete 180° switching of Néel vector. We then reverse the direction of the in-plane magnetic field and the polarity of Hall resistance switching is reversed, agreeing with the SOT switching scheme[29,30].

We further study the temperature- and field-dependence of the switching performance. With the increase of temperature, the switching current can be reduced to 11.4 MA/cm$^2$ at 180 K (Figs. 3d and 3e). Besides, we quantify the SOT efficiency defined as the ratio of effective SOT magnetic field and current density by measuring current-induced hysteresis loop shift (Fig. S7). The SOT efficiency is estimated to be ~38 Oe MA$^{-1}$ cm$^2$, which is remarkably higher than the values reported in typical ferromagnetic systems (*e.g.*, 6.0 Oe MA$^{-1}$ cm$^2$ in Pt/Fe$_3$GaTe$_2$[31] and 7.5 Oe MA$^{-1}$ cm$^2$ in Pt/Co[32]). By measuring the spin-Hall magnetoresistance (SMR) in Pt/FCGT, we reveal that the atomically sharp interface of Pt/FCGT produces a high spin-mixing conductance of $1.93 \times 10^{15}$ $\Omega^{-1}$m$^{-2}$, allowing spin currents efficiently transfer into the adjacent antiferromagnet and thus leading to a high SOT efficiency (Supplementary Note 6). We then vary the magnitude of the in-plane magnetic field, the switching current is almost unchanged (Fig. 3f). For the switching of a ferromagnet, the in-plane magnetic field can significantly tilt the magnetization and change the switching currents. By contrast, the interlayer antiparallel exchange coupling makes FCGT immune to external magnetic fields. We show that by altering the direction of currents, Néel vector can be sequentially switched, showing the reliable switching performance (Fig. 3g).

**Macrospin model for SOT switching in an A-type antiferromagnet**

To explain the mechanism of SOT switching of Néel vector in FCGT, we employ a zero-temperature macrospin model (see Supplementary Note 7) which considers an A-type PMA antiferromagnet with *N* layers. Each antiferromagnet layer experiences current-induced SOTs: $\boldsymbol{H}_{\mathrm{SOT}}^i = H_{\mathrm{SOT}}(\hat{\boldsymbol{m}}_i \times \hat{\boldsymbol{\sigma}})$, where $H_{\mathrm{SOT}}$ and $\hat{\boldsymbol{\sigma}}$ represent the SOT equivalent magnetic field and the direction of spin polarization. We neglect the spin current decay as the spin diffusion length in an antiferromagnet is usually longer than the thickness of our flakes[33,34]. In addition, each antiferromagnet layer experiences the external magnetic field $\boldsymbol{H}_{\mathrm{ext}}$, the effective field induced by PMA: $\boldsymbol{H}_{\mathrm{an}}^i = H_{\mathrm{an}} m_{iz} \hat{\boldsymbol{z}}$, as well as the interlayer exchange coupling: $\boldsymbol{H}_{\mathrm{ex}}^i = -H_{\mathrm{ex}}(\hat{\boldsymbol{m}}_{i-1} + \hat{\boldsymbol{m}}_{i+1})$, where $H_{\mathrm{an}}$ and $H_{\mathrm{ex}}$ represent the magnitudes of anisotropy- and exchange-induced fields, respectively. The equilibrium state of magnetic moment at each layer satisfies the condition of $\hat{m}_i \times (H^i_{SOT} + H_{ext} + H^i_{an} + H^i_{ex}) = 0$, for all $i = 1$ to $N$, and the stability needs be verified by the macrospin dynamics simulation. Note that the equation set of total torque for each layer remain the same when the order of the layer labelling is reversed. Hence, both $\{m_i = m^{eq}_i, i = 1…N\}$ and $\{m_{N+1-i} = m^{eq}_i, i = 1…N\}$ are the equilibrium states, where $\{m^{eq}_i, i = 1…N\}$ is one of the solutions of the equation set. When $N$ is an even number, it indicates that reversing the Néel vector can also satisfy the equilibrium condition, so SOT-induced deterministic switching cannot be achieved in an A-type antiferromagnet with even-number layers.

By solving the condition of total torque equilibrium, we can reproduce the SOT switching behaviour of an A-type antiferromagnet with odd-number layers (Fig. 4a). With a fixed in-plane magnetic field, SOTs gradually rotate the Néel vector within the *yz* plane. Until, for a sufficiently large SOTs, the Néel vector abruptly rotates to the opposite direction, indicating the switching. We next vary the layer number of the antiferromagnet (Fig. 4b). The field-dependence of switching currents in antiferromagnets is very weak, which is different from that in a ferromagnet ($N = 1$) and agrees with the experimental observation (Fig. S6). We further study the spin dynamics during the switching of the Néel vector by solving Landau–Lifshitz–Gilbert (LLG) equations (Supplementary Note 7). As a consequence of strong interlayer exchange interaction, two spin sublattices keep nearly antiparallel and precess together to the opposite direction (Fig. 4c). The strong exchange interaction also yields a fast precession frequency, leading to a fast-switching time (Figs. 4d and S10). Therefore, our theoretical model offers insights into the SOT-induced switching dynamics in A-type antiferromagnets. Note that accurate controlling of atomic layer number is highly challenging in conventional thin-film antiferromagnets. In contrast, vdW antiferromagnets, with the unique atomically exfoliable property, enable precise control over the atomic layer number, thus providing an ideal material platform for SOT switching studies.

**Field-free switching in Pt/FCGT/CrSBr**

In the conventional SOT switching scheme, an in-plane magnetic field is indispensable to achieve the deterministic switching of a PMA magnetization[29,30]. Similar to ferromagnet, we cannot achieve the switching of Néel vector in Pt/FCGT devices in the absence of in-plane magnetic fields (Fig. S12). However, in real applications, this is detrimental for energy consumption and miniaturization considerations. Tremendous efforts have been devoted to eliminate the in-plane magnetic field by introducing mirror-symmetry breaking via spin symmetry engineering[35-38]. Yet, the field-free 180° switching of the Néel vector in an antiferromagnet remains challenging in the experiment.

To break the mirror symmetry, we construct a vdW stack with the noncollinear spin texture by transferring the FCGT flake on the CrSBr flake that is an A-type vdW antiferromagnetic semiconductor (Figs. 5a, 5b and S16)[39]. Particularly, the easy axis of CrSBr is along the *b*-axis, exhibiting a strong in-plane magnetic anisotropy (Fig. S14). The exchange coupling between out-of-plane FCGT and in-plane CrSBr produces an effective in-plane magnetic field on FCGT, causing the tilt of magnetic moments at the FCGT/CrSBr interface along the *b*-axis of CrSBr (Fig. S17). Consequently, we observe a deterministic switching of Hall resistance when the current is applied along the *b*-axis of CrSBr even in the absence of an in-plane magnetic field (Fig. 5c). We further confirm the field-free switching performance by measuring the half loop with current sweeping from zero (Fig. 5d). The device is fully saturated under a preset magnetic field of ±10 kOe. The current-induced Hall resistance switching is ~70%, which is agreed with the switching ratio of as-grown devices shown in Fig. 5c. The partial magnetization switching can be ascribed to the formation of multidomain state in the CrSBr flake[40], and can be improved by reducing the lateral size and thickness of CrSBr flake to facilitate the single-domain state.

To verify the mechanism of field-free switching, we alter the current direction from the *b*-axis to *a*-axis, *i.e.* the hard axis of CrSBr. Then exchange coupling-induced effective in-plane magnetic field is perpendicular to the current, which dissatisfies the deterministic switching scheme. In this case, we do not observe the performance of current-induced Hall resistance switching (Fig. 5e). In addition, we elevate the device temperature to be between the Néel temperatures of CrSBr and FCGT. In this case, the exchange coupling from CrSBr vanishes and current-induced switching of Néel vector is prohibited even when the current is applied along the *b*-axis of CrSBr (Fig. 5f). Moreover, we apply an in-plane magnetic field along the direction of currents. As shown in Fig. 5g, the polarity of current-induced Hall resistance switching is preserved at small magnetic fields of ±100 Oe. Interestingly, with large magnetic fields, the performance of current-induced switching disappears for both positive and negative direction of magnetic fields, due to the occurrence of spin-flop effect in CrSBr (Fig. S15)[39]. At large magnetic fields along the *b*-axis of CrSBr, the spins start to rotate perpendicular to the *b*-axis, leading to the rotating of effective in-plane magnetic fields to be perpendicular to the currents.

**Conclusion and outlook**

Thanks to the atomistic spin engineering capabilities in vdW antiferromagnets, we can unambiguously modulate the intrinsic net magnetic moment parallel to the Néel vector in FCGT by tuning the even/odd parity of vdW layers. We achieve the current-induced 180° switching of Néel vector in the heterostructure of Pt/FCGT with odd-number vdW layers. Combing with theoretical macrospin models, we reveal a layer-dependent parity effect that governs the switching dynamics and establish a universal SOT switching mechanism for collinear antiferromagnets. Moreover, by constructing a heterostructure comprising of CrSBr flake with in-plane easy axis, the interlayer exchange coupling between two antiferromagnets helps to form noncollinear interfacial spin textures and breaks the mirror symmetry, leading to the performance of field-free switching. These results are ascribed to the atomically exfoliable nature of vdW antiferromagnets, which is inaccessible in conventional thin-film antiferromagnets (Supplementary Note 13). Though the Néel temperature of the model vdW antiferromagnet FCGT implemented in our experiments is below room temperature, this paradigm for manipulating antiferromagnetic order is applicable to other vdW antiferromagnets with high Néel temperatures[41-43] and large-scale synthesis capability[44] (Supplementary Note 14). Moreover, the vdW antiferromagnet can exhibit very large tunnelling magnetoresistance, providing an electrical mean of the reliable reading of Néel vector[45-47]. The integration of field-free SOT switching and tunnelling magnetoresistance promise the development of high-performance all-vdW, all-antiferromagnet memory devices.

**Materials and Methods**

**Growth of vdW crystals**

High-quality FCGT single crystals were grown by the self-flux method from a mixture of high purity Fe powders (99.95%), Co powders (99.998%), Ga lumps (99.9999%), and Te powders (99.99%) in a molar ratio of 1.7: 0.3: 1:2. The mixture was sealed in an evacuated quartz tube and heated to 1273 K at a rate of 5 K/min, and then held for one day for solid reactions. Then the temperature was slowly decreased to 1053 K at a rate of 1.8 K/h. Sizable single crystals with a typical size of 2 × 2 × 0.02 mm were obtained after a natural cooling process.

CrSBr single crystals were synthesized by the chemical vapor transport method using chromium (99.99%), sulphur (99.9999%), and bromine (99.9999%), combined with a stoichiometry of 1: 1: 1, sealed in a quartz tube under high vacuum. The quartz tube was heated to 973 K and maintained at this temperature for 15 hours. Then, the source and growth ends were kept at 1123 and 1173 K, respectively. After 25 h, the temperature gradient was reversed and the temperature was maintained for 5 days. The needle-like CrSBr single crystals with the lengths of 2 cm could be obtained after furnace cooling.

**Fabrication of vdW devices**

To prepare the Pt/FCGT devices, FCGT flakes were first transferred onto $SiO_2$/Si substrates using mechanical exfoliation in the glovebox enclosure with inert gas atmosphere. Then, the Pt (6-10 nm) layer was deposited onto the FCGT flake using dc magnetron sputtering (AJA) with a low power of 20 W to eliminate the damage of the FCGT/Pt interface by the bombardment of heavy Pt atoms (Fig. S20). The base pressure for the sputtering was <5 × $10^{-8}$ torr. The Pt layer was further patterned into a Hall bar geometry using ultraviolet (UV) lithography and Ar ion milling, with the FCGT flake located under the cross centre of the Hall bar. The width of the Hall bar is 5-10 μm.

To prepare the Pt/FCGT/CrSBr devices, CrSBr flakes (thickness around 10-30 nm) were mechanically exfoliated on $SiO_2$/Si substrate while FCGT flakes (thickness below 10 nm) were mechanically exfoliated on Polydimethylsiloxane (PDMS) stamps inside glovebox same as up mentioned and sequentially the FCGT flake was staked directly on the top of CrSBr flake with the help of 2D micro transfer system inside glovebox. Then, the Pt (10 nm) layer was deposited onto the bilayer heterostructure using dc magnetron sputtering (AJA) with a low power of 20 W. We then utilized UV lithography and Ar ion milling technique to fabricate Pt/FCGT/CrSBr Hall bar devices. Notably, during the fabrication process, the direction of ***b***-axis in CrSBr is parallel to the direction of two opposing electrodes to fix the direction between pulsed current and CrSBr crystal orientation.

In the experiment, the vdW magnetic material and devices were always kept in the glovebox enclosure with inert gas atmosphere. Before taking out of the glove box, the devices were sealed in a vacuum package to prevent oxidation degradation. The FCGT flakes were exposed to air only during the transferring process from the package to sputtering system. By optimizing the fabrication process, the total transferring time was strictly controlled to be <3 min. Besides, we kept the FCGT flakes in the sample box filled with inert gas during the transferring process, which can further reduce the oxidation degradation.

**Characterization of vdW crystals and devices**

The chemical compositions were identified by the energy dispersive spectroscopy (EDS) measurements at different microregions on the fresh cleavage surface of one single crystal. The crystallographic phase of these single crystals was characterized using room-temperature X-ray diffraction (XRD) measurements (X'Pert Pro MPD diffractometer) with Cu $K\alpha$ radiation ($\lambda = 1.5406$ Å). The magnetization data were measured in a Physical Property Measurement System (PPMS) system.

The characterization of the morphology of the nanoflakes and devices was performed by optical microscopy (Olympus). The cross-section microstructure was investigated using a transmission electron spectroscopy (TEM, JEM-F200, JEM-ARM 200). Devices for TEM were prepared with a focused ion beam system (FIB, TESCAN).

**Transport measurement of vdW devices**

Keithley 2400 source meter and 2182 nanovolt meter were used to apply currents and to collect the Hall resistances, respectively. Specifically, for the SOT switching test, the direction of current pulse was applied along the in-plane magnetic field. A switching current pulse with 1 ms width was applied followed by a relatively small current to read the magnetization state. We define the critical switching current density $J_c = (J_c^+ + J_c^-)/2$, where $J_c^+$ ($J_c^-$) indicates the current density of switching process started (completed). The low temperature magnetotransport properties of the devices were performed in the VSM of a Physical Property Measurement System (DynaCool, Quantum Design) system, which can provide the environment of high three-dimensional vector magnetic fields and varied temperatures. To study the Joule heating effect at high current densities during SOT switching, we conduct the finite element analysis (FEA) simulation and find the temperature increase at the switching current density of 21.8 MA/cm² is ~5 K (see Supplementary Note 12).

**Macrospin simulations**

The macrospin simulations are based on the Landau-Lifshitz-Gilbert (LLG) equation, which includes damping-like SOTs. This can be expressed as: $\frac{d\hat{m}_i}{dt} - \alpha \hat{m}_i \times \frac{d\hat{m}_i}{dt} = -\gamma \hat{m}_i \times H_{\text{eff}}$. In the static simulations, we solve the equilibrium equations: $\hat{m}_i \times H_{\text{eff}} = 0$ using Newton's method. The magnetic parameters, such as the exchange field $H_{\text{ex}} = 1/3 H_{\text{an}}$, are determined from experimental data. A fixed in-plane magnetic field $H_{\text{ext}} = 0.4 H_{\text{an}}$ is applied in the simulations of SOT switching (Fig. 4a).

In the dynamic simulations, we rewrite the LLG equation as: $\frac{d\hat{m}_i}{dt} = -\frac{\gamma}{1+\alpha^2}\left(\hat{m}_i \times H_{\text{eff}} + \alpha \hat{m}_i \times (\hat{m}_i \times H_{\text{eff}})\right)$. We assume that the magnetism of atoms mainly originates from electron spin, so the gyromagnetic ratio $\gamma = 1.76 \times 10^{11}$ T⁻¹ s⁻¹. A Gilbert damping constant of $\alpha = 0.1$ is selected based on the reference literature[48,49]. The Heun method are employed during the simulation[50]. The time step of the simulation is controlled such that the maximum change in $\hat{m}_i$ is less than $1 \times 10^{-2}$. The SOT is introduced at $t = 0$ and persists long enough to induce switching. Additionally, the critical SOT values in the dynamic simulations are given from the static simulations.

**Figures and figure captions**

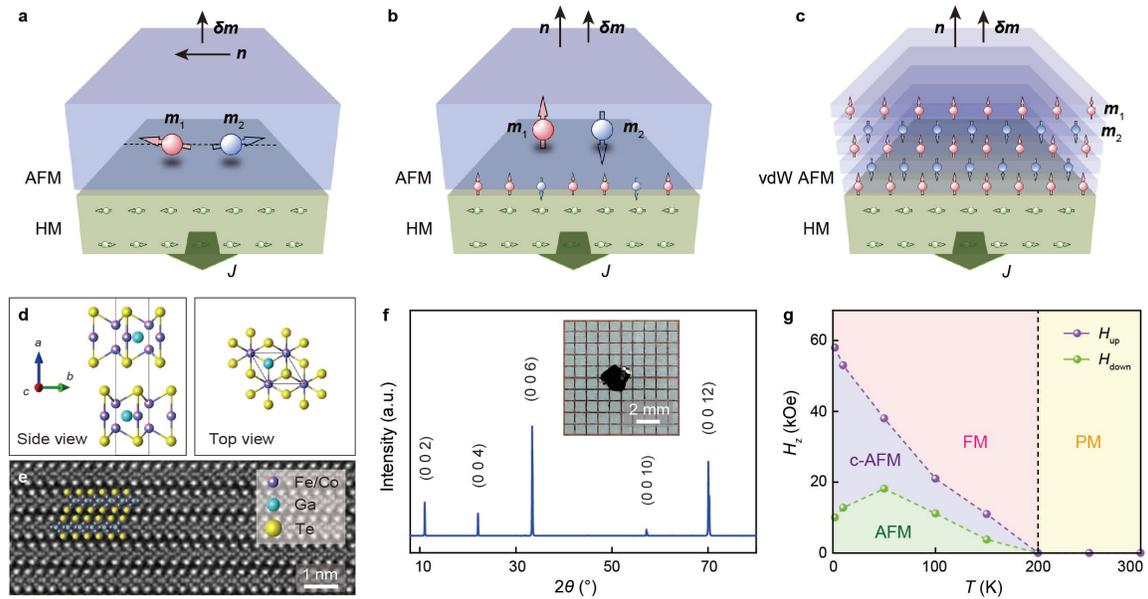

**Figure 1 | Current-induced 180° switching of Néel vector and basic properties of FCGT bulk.** (**a** to **c**) Schematic illustrating current-induced 180° switching of Néel vectors in three types of antiferromagnetic devices comprising of antiferromagnet (AFM)/ heavy metal (HM) bilayers. The net magnet moments **δm** in **a**, **b** and **c**, originate from DMI-induced spin canting, interfacial uncompensated magnetic moment and odd-number vdW layers in an A-type AFM, respectively. The Néel vector is defined as: $\boldsymbol{n} = \boldsymbol{m_1} - \boldsymbol{m_2}$, and the directions of **n** and **δm** are indicated by black arrows. The green spheres with arrows denote the current-induced electron spin accumulation in HM, while the red and blue spheres with arrows denote two spin sub-lattices in AFMs. (**d**) Crystal structure of FCGT viewed from the top and the side. (**e**) High resolution transmission electron microscopy (HR-TEM) image of FCGT bulk. The elemental atoms are indicated by yellow (Te), blue (Ga) and purple (Fe/Co) circles. (**f**) X-ray diffraction (XRD) pattern of FCGT bulk and the optical image (inset) of FCGT single crystal. (**g**) Magnetic phase diagram of FCGT bulk with applied magnetic field $H_z$ along ***c***-axis. The switching fields for upward ($H_{up}$) and downward ($H_{down}$) sweeping define the magnetic phases of paramagnet (PM), ferromagnet (FM), canted AFM (c-AFM) and AFM.

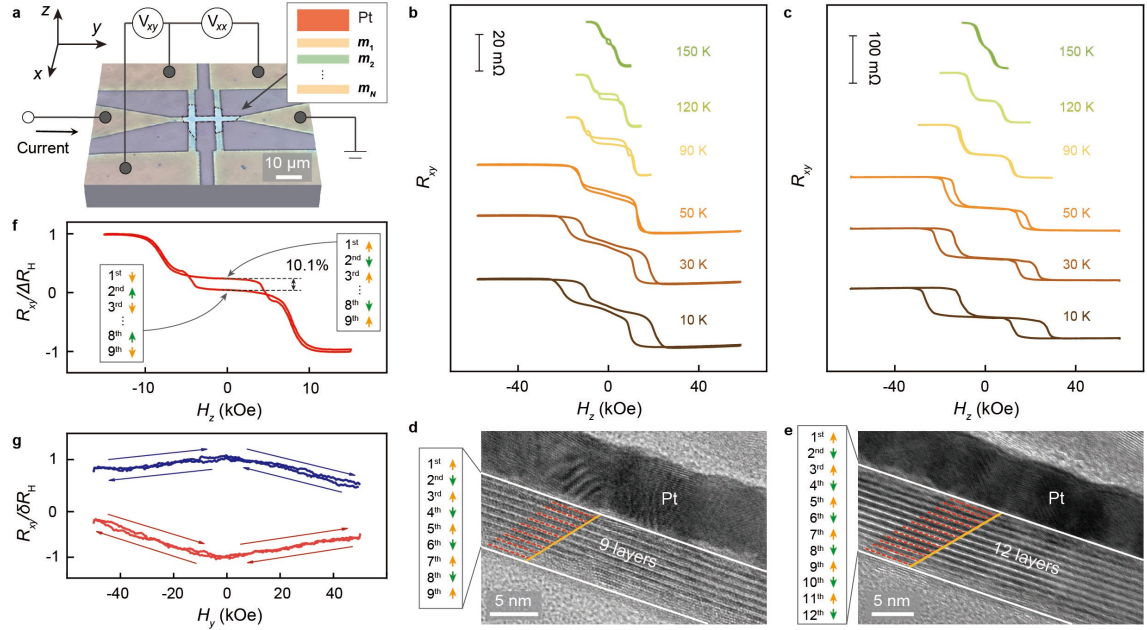

**Figure 2 | Intrinsic magnetic moment in Pt/FCGT with odd-number vdW layers.** (**a**) Coloured optical microscope image of Pt/FCGT device in a three-dimensional rendering of the transport measurement setup. The area of FCGT flake is indicated by black dashed lines. (**b** and **c**) Temperature-dependent AHE hysteresis loops of Pt/FCGT devices with odd- (**b**) and even- (**c**) number vdW layers. (**d** and **e**) Cross-sectional HR-TEM images of Pt/FCGT devices with odd- (**d**) and even- (**e**) number vdW layers. The white lines indicate the interfaces of FCGT flakes and red dashed lines indicate the vdW layers. (**f**) AHE hysteresis loop of 9-layers FCGT at 120 K. (**g**) AHE of 9-layers FCGT with the preset of out-of-plane magnetic fields of 20 kOe (red line) and -20 kOe (blue line) by applying in-plane magnetic fields at 120 K. The arrows indicate the directions of field sweeping.

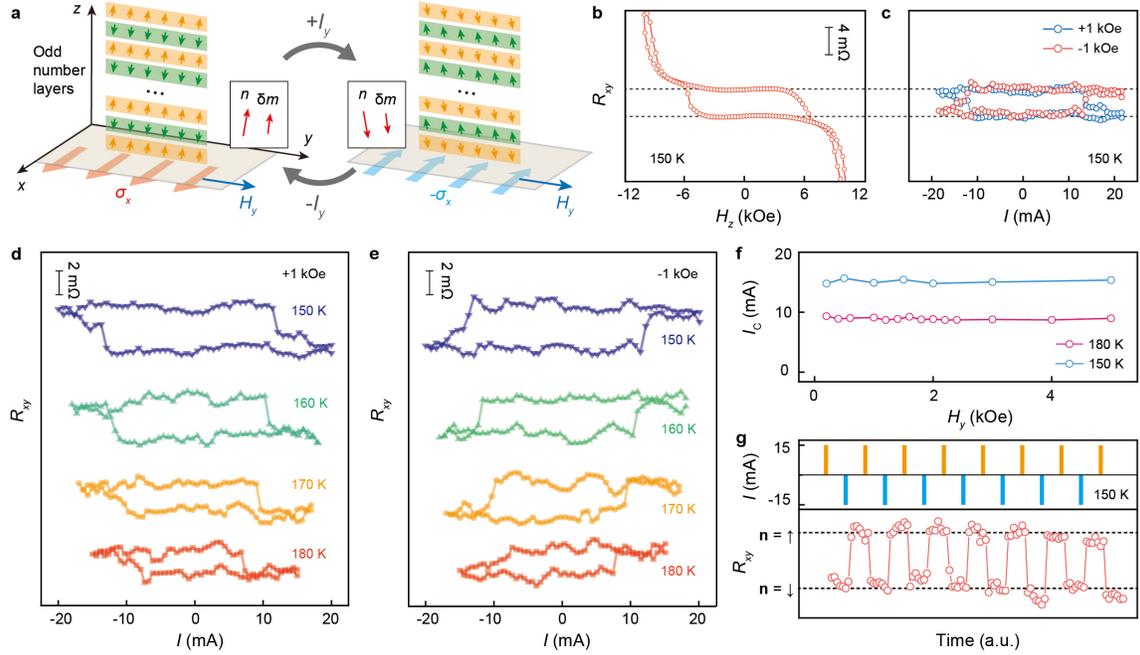

**Figure 3 | Current-induced switching of Néel vector in Pt/FCGT.** (**a**) Schematics illustrating SOT 180° switching of the Néel vector in Pt/FCGT with odd-number vdW layers. The directions of applied current $I_y$, spin polarization $\sigma$, in-plane magnetic field $H_y$, Néel vector $\boldsymbol{n}$ and net magnetic moment $\boldsymbol{\delta m}$ are indicated. (**b**) AHE hysteresis loop of $R_{xy}$ driven by out-of-plane magnetic fields $H_z$. (**c**) Current-induced AHE hysteresis loops with in-plane magnetic fields of ±1 kOe. The dashed lines in (**b**) and (**c**) indicate the magnitude of $\delta R_H$, showing the nearly full switching of Néel vector. (**d** and **e**) Temperature-dependent current-induced AHE hysteresis loops with the in-plane magnetic fields of 1 kOe (**d**) and –1 kOe (**e**). (**f**) Switching current as a function of in-plane magnetic fields at 180 K and 150 K. (**g**) Robust 180° switching of Néel vector with alternating currents with in-plane magnetic fields of 1 kOe at 150 K. Top panel: Sequence of alternating currents; bottom panel: corresponding measured AHE resistance indicating the direction Néel vector.

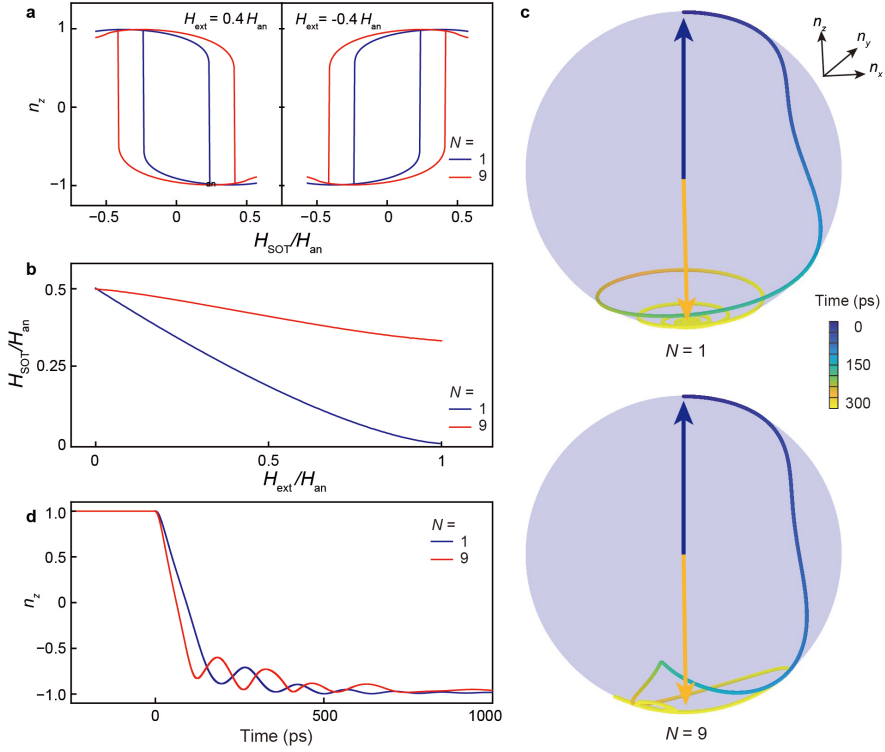

**Figure 4 | Macrospin model of SOT switching in an A-type antiferromagnet.** (**a**) Simulated curves of SOT 180° switching of the Néel vector in FCGT of 1 and 9 layers at fixed in-plane magnetic fields of ±0.4$H_{an}$. (**b**) Magnitude of SOTs to switch the Néel vector as a function of in-plane magnetic fields by macrospin simulations. (**c**) Trajectories of Néel vector during the switching by macrospin simulations in FCGT with 1 and 9 layers. The colour in trajectory lines indicates the time. (**d**) Simulated dynamics of out-of-plane Néel vector in FCGT with 1 and 9 layers. The in-plane magnetic field of 0.4$H_{an}$ and the critical switching SOTs are applied at $t = 0$. For $N = 1$, the direction of ***n*** is given by ***m***$_1$.

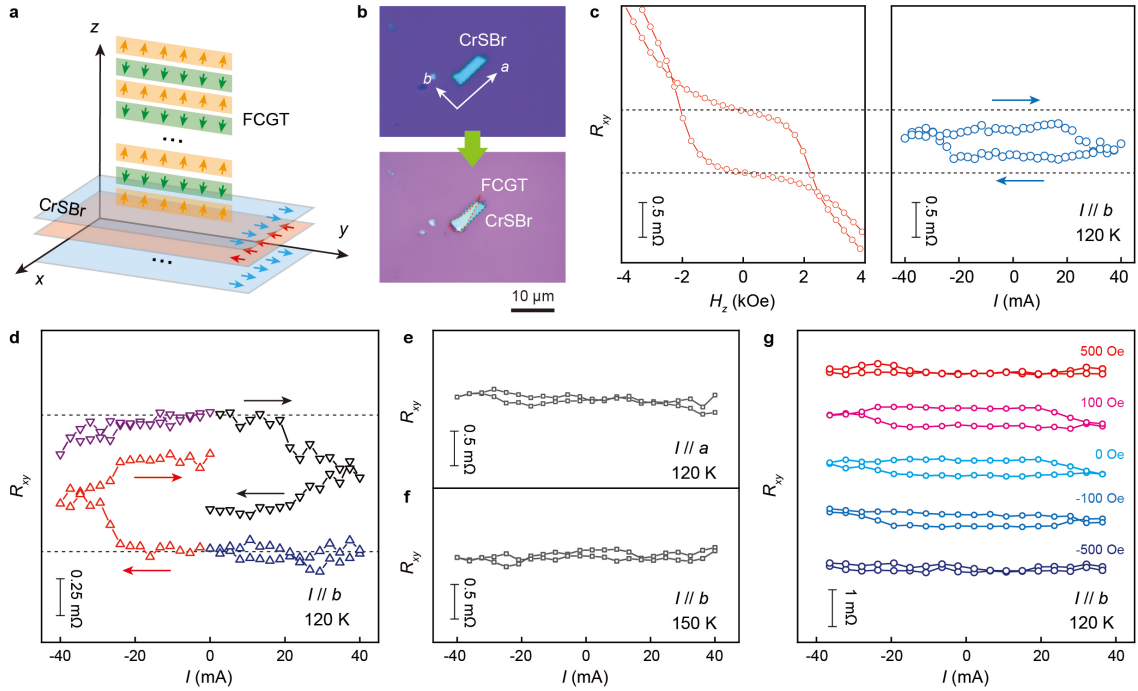

**Figure 5 | Field-free switching of Néel vector in Pt/FCGT/CrSBr.** (**a**) Schematics illustrating field-free 180° switching of the Néel vector in Pt/FCGT/CrSBr heterostructures. (**b**) Optical images of CrSBr flake and FCGT/CrSBr heterostructure. Green and red dashed lines show the area the CrSBr and FCGT flakes. The directions of the *a*- and *b*-axis of the CrSBr flake are indicated. (**c**) AHE hysteresis loops driven by out-of-plane magnetic fields (left panel) and currents applied along the direction of *b*-axis in CrSBr (right panel). The direction of current-induced switching is indicated. (**d**) Four half-loops driven by currents applied along the direction of *b*-axis in CrSBr. The device was preset by out-of-plane magnetic fields of of 10 kOe (red and blue triangles) and -20 kOe (purple and black triangles). Arrows indicate the direction of current sweeping. (**e**) AHE hysteresis loops driven by currents applied along the direction of *a*-axis in CrSBr. (**f**) AHE hysteresis loops driven by currents applied along the direction of *b*-axis in CrSBr at 150 K above the Néel temperature of CrSBr. (**g**) AHE hysteresis loops driven by currents applied along the direction of *b*-axis in CrSBr with various in-plane magnetic fields.